\documentclass[showpacs,prb,twocolumn]{revtex4}
\usepackage{amsfonts}
\usepackage{amssymb}
\usepackage{amsmath}
\usepackage{graphicx}


\begin{document}

\title{Conditions for magnetically induced singlet d-wave superconductivity on the square lattice}

\author{S.~R.~Hassan $^{1}$, B.~Davoudi$^{1,2}$, B. Kyung $^{1}$ and A.-M.S.~Tremblay$^{1}$}
\begin{abstract}
It is expected that at weak to intermediate coupling, d-wave superconductivity can be induced by antiferromagnetic fluctuations. However, one needs to clarify the role of Fermi surface topology, density of states, pseudogap, and wave vector of the magnetic fluctuations on the nature and strength of the induced d-wave state. To this end, we study the generalized phase
diagram of the two-dimensional half-filled Hubbard model as a function of
interaction strength $U/t$, frustration induced by second-order hopping $t^{\prime }/t$, and temperature $T/t$. In experiment, $U/t$ and $t^{\prime }/t$ can be controlled by pressure. We use the two-particle self-consistent approach (TPSC), valid from weak
to intermediate coupling. We first calculate as a function of $t^{\prime}/t$ and $U/t$ the temperature and wave vector at which the spin response function begins to grow exponentially.
D-wave superconductivity in a half-filled band can be induced by such magnetic fluctuations at weak to intermediate coupling,but only if they are near commensurate wave vectors and not too close to perfect nesting conditions where the pseudogap becomes detrimental to superconductivity. For given $U/t$ there is thus an optimal value of frustration $t^{\prime}/t$ where the superconducting $T_c$ is maximum. The non-interacting density of states plays little role. The symmetry d$_{x^{2}-y^{2}}$ vs d$_{xy}$ of the superconducting order parameter depends on the wave vector of the underlying magnetic fluctuations in a way that can be understood qualitatively from simple arguments.


\end{abstract}

\affiliation{$^{1}$D\'{e}partment de Physique and RQMP, Universit\'{e} de Sherbrooke,
Sherbrooke, Qu\'{e}bec, Canada J1K 2R1\\
$^{2}$Institute for Studies in Theoretical Physics and Mathematics, Tehran
19395-5531, Iran}
\maketitle

\section{Introduction}

Even before the discovery of high-temperature superconductivity, research in organic and heavy fermion compounds lead to the suggestion that antiferromagnetic spin fluctuations can promote  superconductivity with an order parameter that changes sign along the diagonal of the Brillouin zone, so-called d-wave superconductivity~\cite{Beal-Monod:1986,Caron:1986,Scalapino:1986,Miyake:1986}. This question can be studied using the Hubbard model. At strong coupling, namely interaction strength $U$ larger than the bandwidth, mean-field factorization of the equivalent $t-J$ model already reveals the possibility of d-wave pairing. The situation is quite different at weak coupling. There, no mean-field factorization of the Hubbard model leads to a d-wave superconducting state. Instead, magnetic fluctuations are dominant. Pairing must then be seen in a two-stage process. First, sufficiently strong magnetic fluctuations are formed, then the corresponding low energy bosons can act as a glue for pairs, in a manner analogous to what phonons do in the standard BCS theory. In other words, at strong coupling it seems that superexchange suffices to form pairs, while at weak coupling there is an intermediate process where mediating bosons must be formed before they can bind pairs~\cite{Anderson:2007}.

There is now ample numerical and analytical evidence for d-wave superconductivity in the one-band Hubbard model~\cite{Bickers_dwave:1989,Scalapino:1995,Carbotte:1999,Monthoux:1999,Moriya:2003,Senechal:2005,Maier_d:2005,LTP:2006,Bourbonnais:2006,Haule:2007,Kancharla:2007,Reiss:2007,Sorella:2007,Gros:2007,Lee:2007} even though there are some dissenting voices~\cite{Imada:2007}. While the coupling in high-temperature superconductors or the layered BEDT organics is strong enough to lead to a Mott insulating state at half-filling, it is still interesting to understand more deeply the physics at weak to intermediate coupling in the absence of the Mott insulating state. There, numerical methods are limited by the fact that small system sizes do not allow magnetic fluctuations at incommensurate wave vectors.

In this paper, we study the two-dimensional Hubbard model at half-filling. In two-dimensions, the Mermin-Wagner theorem implies that there is no long-range magnetic order at finite temperature. Nevertheless, there is a crossover temperature at which the magnetic correlation length starts to grow exponentially as temperature decreases. One enters the so-called renormalized classical regime. We thus study the influence of second-neighbor hopping $t^\prime$ and of interaction strength $U$ on the magnetic crossover temperature. Berezinsky-Kosterlitz-Thouless superconductivity is allowed in two-dimensions and can be induced by these magnetic fluctuations. We answer the question of the relation between the wave vector of the dominant magnetic fluctuations and the symmetry and magnitude of d-wave superconductivity~\cite{Powell:2007}, as well as the question of the relative importance of the non-interacting density of states, the double occupancy, the magnetic correlation length, and the detailed shape of the Fermi surface on superconductivity.

We will use the Two-particle self-consistent (TPSC) approach~\cite{Vilk:1994,Vilk:1997,Allen:2003,Kyung:2003} that is non-perturbative and valid at weak to intermediate coupling. This method has been benchmarked by comparisons with Quantum Monte Carlo calculations wherever possible. It gives results for the pseudogap that are in agreement with experiments in electron-underdoped cuprates~\cite{Kyung:2004,Motoyama:2007}. In addition, the superconducting transition temperature obtained~\cite{Kyung:2003} by this method is very close to more accurate results from the Dynamical Cluster Approximation~\cite{Maier_d:2005}.

      We first introduce the model and the TPSC method, explaining in detail why it is a method of choice for this problem. We next present our  numerical
      results, discuss  various physical aspects that can be
      found  in the $t-t^\prime-U$ model and compare our results with other results in earlier literature. We then discuss the relationship between d-wave superconductivity and magnetic fluctuations in terms of a simple physical picture and conclude.

\section{Model And Formalism}
\label{sec2}

We study the Hubbard Hamiltonian,
\begin{align}
H  & =-t\sum_{\langle i,j\rangle,\sigma}\left(  c_{i\sigma}^{\dagger
}c_{j\sigma}+h.c.\right)  -t^{\prime}\sum_{\left\langle \langle i,j\rangle
\right\rangle ,\sigma}\left(  c_{i\sigma}^{\dagger}c_{j\sigma}+h.c.\right)
\nonumber\\
& +U\sum_{i}n_{i\uparrow}n_{i\downarrow}%
\end{align}
where $c_{\mathbf{i}\sigma }$ ($c_{\mathbf{i}\sigma }^{\dagger }$) are annihilation (creation) operator for electrons of spin $\sigma $ at site $i$. The hopping matrix element $t$ is between nearest-neighbors $<ij>$ whereas $t^\prime$ is between next-nearest-neighbors $<<ij>>$, $n_{\mathbf{i}\sigma }$ is the density operator at site $i$, and $U$ is the on-site interaction. Particle-hole transformation with and without a phase factor shows that at half-filling the results do not depend on the sign of either $t$ or $t^\prime$.

The TPSC approach that we use in this paper satisfies the Pauli principle $<n_{\sigma}^2>=<n_{\sigma}>$, conservation of spin and charge as well as the Mermin-Wagner theorem. In addition, it includes renormalization of interaction (Kanamori-Br\"uckner) and has a self-energy that includes the vertex corrections that must be taken into account in the absence of the conditions necessary to satisfy Midgal's theorem. TPSC also achieves consistency between one- and two-particle quantities in the form of the sum rule relating potential energy to the trace of self-energy times Green function. At weak to intermediate coupling ($U \lesssim 6t$), detailed comparisons with benchmark Quantum Monte Carlo simulations~\cite{Vilk:1997,LTP:2006,Kyung:2003} reveal that TPSC is accurate to a few percent.

In this approach, spin and charge susceptiblities $\chi_{sp}$, $\chi_{ch}$ have random phase approximation (RPA)-like
forms but with two different effective interactions $U_{sp}$ and $U_{ch}$ that are then determined self-consistently. Although the susceptiblities have a RPA functional
form, the physical properties of the theory are very different from the RPA because of the self-consistency condition on $U_{sp}$ and $U_{ch}$. The necessity to have
two different effective interactions for spin and charge is dictated by the Pauli exclusion principle $<n_{\sigma}^2>=<n_{\sigma}>$ which implies that the local values of $\chi_{sp}$
and $\chi_{ch}$ are related to only one local pair correlation function $<n_{\uparrow}n_{\downarrow}>$. Indeed, using the fluctuation-disspation theorem in Matsubara
formalism, we have the exact sum rules
\begin{equation}
<n_{\uparrow}^2>+<n_{\downarrow}^2>+2<n_{\uparrow}n_{\downarrow}>-n^2=\frac{1}{\beta N}\sum_{\tilde q}\chi_{ch}({\tilde q})
\end{equation}
and
\begin{equation}
<n_{\uparrow}^2>+<n_{\downarrow}^2>-2<n_{\uparrow}n_{\downarrow}>=\frac{1}{\beta N}\sum_{\tilde q}\chi_{sp}({\tilde q}),
\end{equation}
where $\beta\equiv\frac{1}{T}, n=<n_{\uparrow}>+<n_{\downarrow}>, {\tilde q}=({\bf q}, iq_{n})$ with ${\bf q}$ the wave
vectors of an N-site lattice, and with $iq_{n}=2\pi i n T$ the bosonic Matsubara frequencies. The Pauli principle $<n_{\sigma}^2>=<n_{\sigma}>$ applied to the
left-hand side of both equations with the TPSC expressions for $\chi_{sp}$, $\chi_{ch}$ on the right-hand side leads to

\begin{equation}
n+2<n_{\uparrow}n_{\downarrow}>-n^2=\frac{T}{N}\sum_{\tilde q}\frac{\chi_{0}({\tilde q})}{1+\frac{1}{2}U_{ch}\chi_{0}({\tilde q})},
\end{equation}

\begin{equation}
n-2<n_{\uparrow}n_{\downarrow}>=\frac{T}{N}\sum_{\tilde q}\frac{\chi_{0}({\tilde q})}{1-\frac{1}{2}U_{sp}\chi_{0}({\tilde q})},
\end{equation}
with $\chi_{0}({\tilde q})$ the susceptiblities calculated from Green functions whose constant self-energy can be absorbed in the chemical potential. Hence, $\chi_{0}({\tilde q})$ takes the non-interacting form.

If $<n_{\uparrow}n_{\downarrow}>$ is known, $U_{sp}$ and $U_{ch}$ are determined from the above equations. The key quantity  $<n_{\uparrow}n_{\downarrow}>$
can be obtained from Monto Carlo simulations or by other means. However, it can also be obtained self-consistently by adding to above set of equations the
relation
\begin{equation}
U_{sp}=g_{{\uparrow}{\downarrow}}(0)U, \, \, \, \, \,  g_{{\uparrow}{\downarrow}}(0)=\frac{<n_{\uparrow}n_{\downarrow}>}{<n_{\uparrow}><n_{\downarrow}>}.
\end{equation}

Equations (5) and (6) define a set of self-consistent equations for $U_{sp}$ that involve
only two-particle quantities. $U_{ch}$ is determined by substituting in Eq.(4) the value of $<n_{\uparrow}n_{\downarrow}>$ that is obtained from Eqs. (5) and (6).

Following the procedure introduced in Ref.~\onlinecite{Moukouri:2000,Allen:2003} one can show that the self-energy can be calculated from
\begin{eqnarray}
\Sigma^{\left( 2\right) } _{\sigma }(\tilde{k})&=&  ( Un_{\tilde{\sigma}})
+\frac{TU}{8N}\sum_{\tilde q}\{3U_{sp}\chi_{sp}(\tilde{q})\nonumber \\
&&+U_{ch}\chi _{ch}(\tilde{q})\}G^{\left( 1\right) }(\tilde{k}+\tilde{q})
\label{Sigmak}
\end{eqnarray}
where the Matsubara frequency associated with $\mathbf{k}$ is fermionic and that associated with $\mathbf{q}$ is bosonic. In $G^{\left( 1\right)}$ the self-energy is a constant that can be absorbed in the chemical potential so that effectively $G^{\left( 1\right)}$ is the non-interacting Green function for half-filling.

As usual we can compute the spectral function from
$A({\mathbf k},\omega)=-Im G^{\left( 2\right) }({\mathbf k},\omega)/\pi$ where
\begin{equation}
G^{\left( 2\right)R }({\mathbf k},\omega)
=\frac{1}{\omega+i\delta-\epsilon_{\bf k}+\mu-\Sigma^{R}(\bf k,\omega)}.
\end{equation}
The interacting chemical potential $\mu$ is found by fixing the number of particles calculated from $G^{\left( 2\right)R }$ to half-filling.

The above formalism can be extended~\cite{Kyung:2003} to compute pairing
correlations. Basically, the above steps are repeated in the presence
of an infinitesimal external pairing field that is eventually set to
zero at the end of the calculation. This allows us to obtain the
particle-particle irreducible vertex in Nambu space from the
functional derivative of the off-diagonal $\Sigma ^{\left( 2\right) }$
with respect to the off-diagonal Green function. The $d$-wave
susceptibility is defined by $\chi _{d}=\int_{0}^{\beta }d\tau
\left\langle T_{\tau }\Delta \left( \tau \right) \Delta ^{\dagger
}\right\rangle $ with the $d$-wave pair creation operator $\Delta ^{\dagger
}=\sum_{i}\sum_{\gamma }g\left( \gamma \right) c_{i\uparrow }^{\dagger
}c_{i+\gamma \downarrow }^{\dagger }$, $g\left( \gamma \right)$ being
a form factor for a representative gap symmetry,
$\beta\equiv 1/T$, $T_{\tau }$ is the time-ordering operator, and $\tau $ is
imaginary time. The final expression for the d-wave
susceptibility in the zero-frequency limit can be written as follows, if we use the notation where $\tilde{k}$ and $\tilde{k}^\prime$ denote both wave vector and fermionic Matsubara frequency,
\begin{widetext}
\begin{align}
\chi _{d}\left( \tilde{q}=0 \right) & =\frac{T}{N}\sum_{\tilde{k}}\left(
g_{d}^{2}({\bf k}) G_{\uparrow }^{\left( 2\right) }\left( -\tilde{k}\right)
G_{\downarrow }^{\left( 2\right) }\left( \tilde{k}\right) \right) -\frac{U}{4}\left(
\frac{T}{N}\right) ^{2}\sum_{\tilde{k},\tilde{k}^{\prime }}g_{d}( {\bf k}) G_{\uparrow
}^{\left( 2\right) }\left( -\tilde{k}\right) G_{\downarrow }^{\left( 2\right)
}\left( \tilde{k}\right)  \notag \\
& \times \left( \frac{3}{1-\frac{U_{sp}}{2}\chi _{0}\left( \tilde{k}^{\prime
}-\tilde{k}\right) }+\frac{1}{1+\frac{U_{ch}}{2}\chi _{0}\left( \tilde{k}^{\prime }-\tilde{k}\right)
}\right) G_{\uparrow }^{\left( 1\right) }\left( -\tilde{k}^{\prime }\right)
G_{\downarrow }^{\left( 1\right) }\left( \tilde{k}^{\prime }\right) g_{d}({\bf
k^{\prime}}).
\label{Suscep_d}
\end{align}
\end{widetext}
In this equation, $g_{d}(\mathbf{k})=\left( \cos k_{x}-\cos k_{y}\right) $
for $d_{x^2-y^2}$ symmetry and $g_{d}(\mathbf{k})=\sin k_{x} \sin k_{y}$ for $d_{xy}$ symmetry.
Details of the calculational procedure are explained in Ref.~\onlinecite{Kyung:2003}. The only difference is that, in the present paper, double occupancy is always computed self-consistently with Eqs.(5) and (6). At high temperature the pairing susceptibility obtained from the above approach is in quantitative agreement with benchmark Quantum Monte Carlo simulations~\cite{Kyung:2003}. However, because of the sign problem, the latter are not available at very low temperatures. Nevertheless, temperatures in the simulations are low enough to observe the effect of the pseudogap.

Since the expression for the susceptibility Eq.~(\ref{Suscep_d}) contains only the
first two terms of the infinite Bethe-Salpeter series, we use
the temperature where the contribution of the
vertex part (exchange of one spin and charge fluctuation) becomes
equal to that of the direct part of the $d$-wave pairing
susceptibility~\cite{Kyung:2003} as a
rough estimate for the transition temperature for $d$-wave
superconductivity. In other words, we look for the equality
of the sign and the magnitude of the two terms appearing in
Eq.~(\ref{Suscep_d}). This choice is motivated by the statement that
$1/\left( 1-x\right) \sim 1+x$ diverges when $x=1$. The true
Kosterlitz-Thouless transition temperature in two dimensions is
expected to occur below the temperature determined from the
Bethe-Salpeter equation. The temperatures that we will find for the transition temperature are thus upper bounds.

Strictly speaking, TPSC looses its quantitative accuracy when it is taken below the crossover to the renormalized classical regime. In cases where we will find the superconducting $T_c$ far below the crossover to the renormalized classical regime, we expect our results to be only qualitatively correct. Nevertheless,
as a further check of the validity of this approach, we point out that detailed finite size studies near optimal doping ($x=0.10$) using the dynamical cluster approximation~\cite{Maier_d:2005} find that $T_c=0.023t$ when
$U=4t, t^\prime=0$ while with TPSC~\cite{Kyung:2003} the corresponding value is $T_c=0.029t$.

\begin{figure}[tbp]
\begin{center}
\includegraphics[scale=0.325]{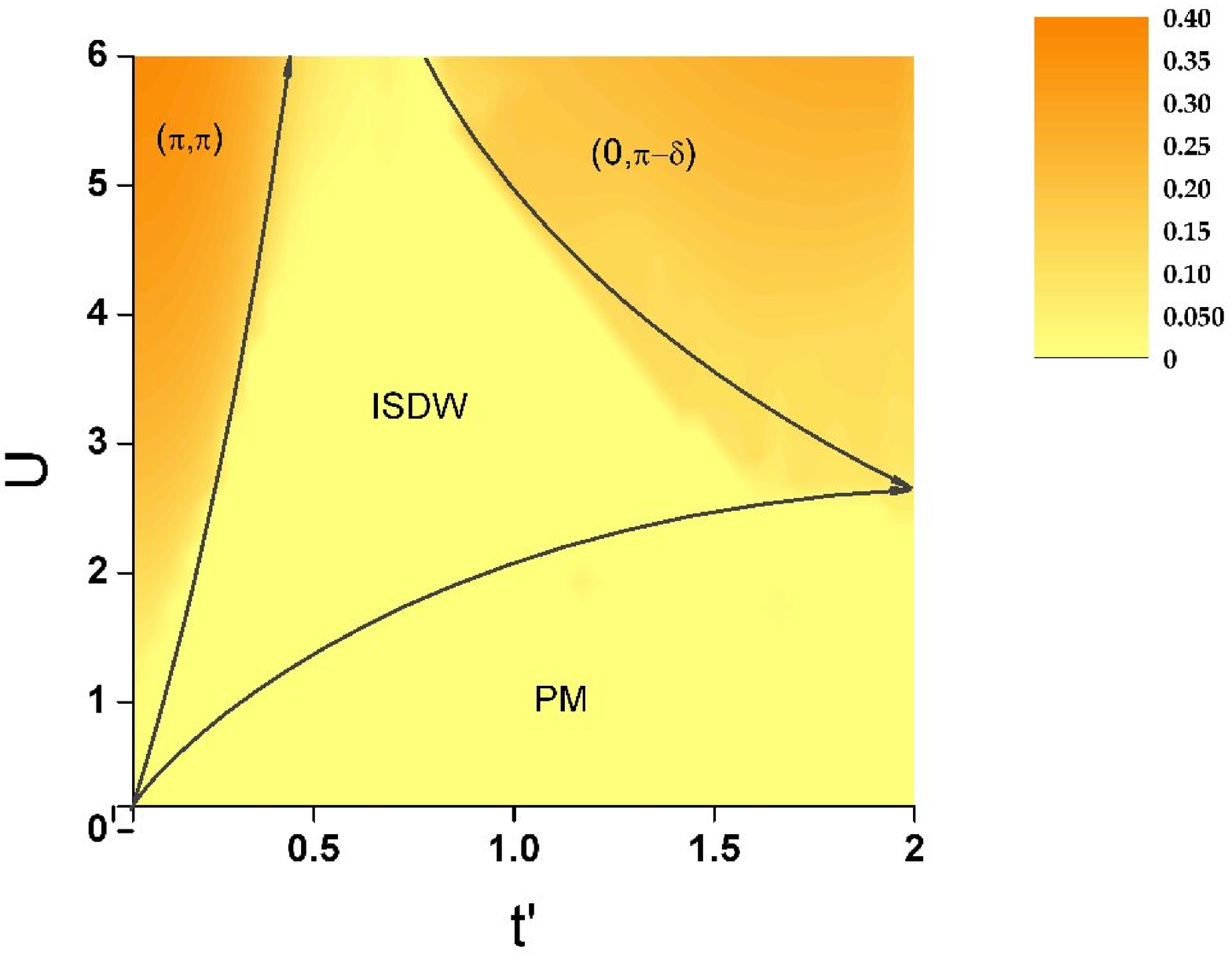}
\includegraphics[scale=0.325]{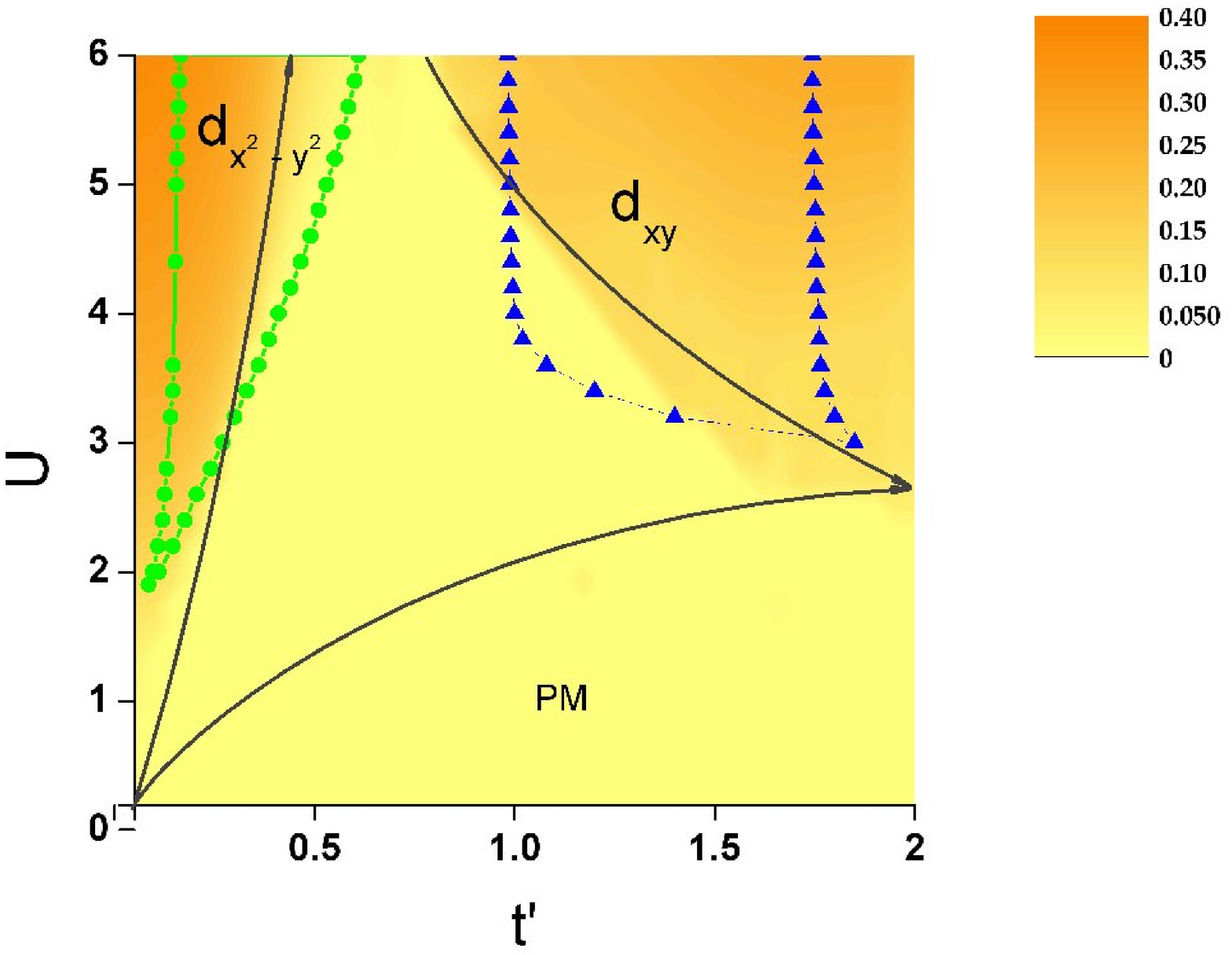}
\end{center}
\caption{(Color online) Crossover diagram at $n=1$ in the $U-t^\prime$ plane, where $U$ and $t^\prime$ are measured in units of hopping $t$. In the top panel there is no superconductivity. In the bottom panel, the boundaries where d-wave superconductivity can appear are indicated. The color scale represents the value
of the crossover temperature for the spin response function. ISDW refers to incommensurate spin-density wave and PM to paramagnetic phase.}
\label{Phase_Diagram}
\end{figure}

\section{Numerical Results and comparisons with other approaches}

 Let us first consider magnetic properties. In two dimensions, there is no long-range magnetic order at finite temperature because of the Mermin-Wagner theorem. Nevertheless, there can be a crossover temperature $T_{X}$ where the antiferromagnetic correlation length ($\xi$) begins
 to increase exponentially. In such a case, $\xi$ becomes infinite only at zero temperature where long-range order is established. One can define the crossover temperature $T_{X}$ as the temperature at which the antiferromagnetic correlation length $\xi$ becomes equal to the thermal de Broglie wavelength. At this temperature, a pseudogap begins to appear at hot spots in the zero frequency single-particle spectral weight~\cite{Vilk:1997}. Hot spots are points of the Fermi surface that are linked by the wave vector where the magnetic fluctuations are becoming large.

In practice,
 as soon as the correlation length becomes relatively large compared with the lattice spacing,
 the rise in $\xi$ as temperature is lowered is rather sharp. For convenience, We thus choose $T_{X}$ as that
 temperature where  $\chi_{sp}(q_{x},q_{y},0)/\chi_{0}(q_{x},q_{y},0)=200$, with  $q_{x}$ and $q_{y}$
the wave vectors where the response function has a maximum. We chose $200$
  for the sake of computational time and checked that the result did not
  change appreciably if we chose a higher value.

 In the top panel of Fig.~\ref{Phase_Diagram} we show the crossover diagram
 of the frustrated Hubbard model on the square lattice in the $U-t^\prime$ plane without superconductivity. All parameters with dimension of energy are measured in units of nearest-neighbor hopping $t$. Boltzmann's constant is also taken as unity.
  The color code
  in Fig.1 represents the variation  of the crossover  temperature.
  The areas indicated with $(\pi,\pi)$, ISDW, $(0,\pi-\delta$),
  and PM are  the commensurate, incommensurate, incommensurate close
  to $(0,\pi)$ ($\delta$ approaches 0 as $t^\prime$ increases in that
  region (see Fig.2)) and nonmagnetic (paramagnetic) metallic state, respectively. The PM state
  indicates that the peak  of the spin response function
  does not grow even at temperature as low as $T=0.01$.

  We note that at small $t^\prime$, the largest peak value of the spin
  response function is at the $(\pi,\pi)$ wave vector,
  which is expected to mediate $d_{{x^2}-y^2}$ superconductivity. For  $t^\prime>0.8$ on the other hand, the response is peaked at the $(0,\pi-\delta)$ wave vector, expected to be relevant for $d_{xy}$ superconductivity.
  Due to the exponentially increasing correlation length with
  decreasing temperature, a small coupling  between two dimensional planes
  would lead to long range spin order at finite temperature, especially for
  small $t^\prime$ and $t^\prime>0.8$.

  Our crossover diagram (without superconductivity) can be compared with zero-temperature results obtained by other methods. For $t^\prime\le 0.5$ there are results of the optimized variational
Monte Carlo (VMC) method in Ref.~\onlinecite{Yokoyama:2006}. For $U<6$, their antiferromagnetic region extends only up to
$|t^\prime/t| \lesssim 0.2$, a result quite different from other studies including ours. For $t^\prime\le 1$ the phase diagram in the $U-t^\prime$ plane
  has been obtained by the path-integeral renormalization group (PIRG) approach~\cite{Mizusaki:2006}. Restricting ourselves to the region
$U < 6$ of that paper, we find that the area of our PM state is smaller. However, the region where $(\pi,\pi)$ fluctuations occur is quite close. They find a $(\pi,0)$ phase for $t^\prime \gtrsim 0.7$ like us but at $U \gtrsim 8$ Due to the finite size of the PIRG calculations, they are not sensitive to the incommensurate spin density waves that we find. That may explain why their paramagnetic region is larger. Also, in the region where we find ISDW they find, at large enough $U$, a non-magnetic Mott insulator (NMI) and at even larger $U$ a $(\pi/2,\pi)$ phase. Our TPSC approach cannot describe the Mott transition, so the question is whether that transition can occur at values as small as about $U=4$ as found in Ref.~\onlinecite{Mizusaki:2006}. That seems to be possible within the Variational
  cluster approximation (VCA) \cite{Andriy:2007}, but VCA overestimates the effect of interactions. The $(\pi,\pi)$ region in the latter work is larger than ours, again perhaps because the small system sizes are not sensitive to incommensurate density waves. However, the $(0,\pi-\delta)$ region is quite close to ours. We should also keep in mind that we are quoting the position of the maximum spin susceptibility at the crossover temperature. That position may change when zero temperature is reached. In fact, quite generally, it is expected that the magnetic fluctuations become more incommensurate as temperature decreases.

\begin{figure}[tbp]
\begin{center}
\includegraphics[scale=0.4]{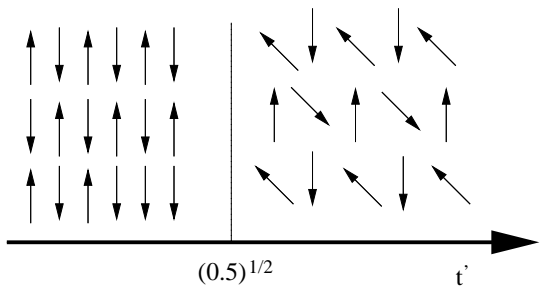}
\end{center}
\caption{The classical ground state~\cite{Misguich:2005} of $J_{1}-J_{2}$ Heisenberg model at T=0 as a function of $t^{\prime}$ obtained by minimizing the dispersion relation
$J({\bf q})=2[cos(q_{x})+cos(q_{y})]+2{t^\prime}^2[cos(q_{x}+q_{y})+cos(q_{x}-q_{y})]$.}
\label{ClassicalGS}
\end{figure}

\begin{figure}[tbp]
\begin{center}
\includegraphics[scale=0.3]{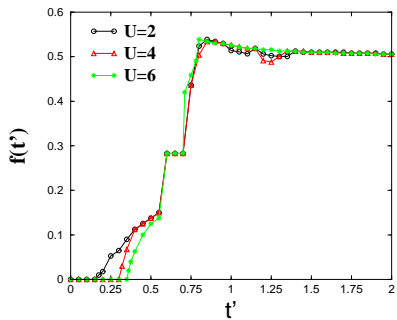}
\end{center}
\caption{(Color online) $f(t^\prime)=\sqrt{(0.5-k_x^{max})^2+(0.5-k_y^{max})^2}$ as a measure of the typical position of the wave vector where the maximum spin response function is located at $T_X$ as a function of $t^\prime$ for $U=2$, $4$ and $6$ all at $n=1$. All energies are in units of $t$.}
\label{f(t)}
\end{figure}

\begin{figure}[tbp]
\begin{center}
\includegraphics[scale=0.3]{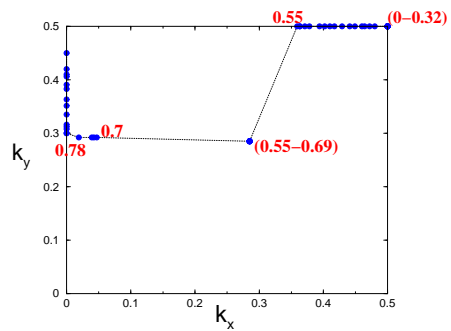}
\end{center}
\caption{(Color online) The position of the wave vector where the maximum spin fluctuations occur in the Brillouin zone for $n=1, U=4$ as a function of $t^\prime$. Numbers near data points indicate the value of $t^\prime$.}
\label{Position-q}
\end{figure}

We open a parenthesis on the incommensurability discussed above. It can be contrasted with what is found in spin models that correspond to the large $U$ limit of the Hubbard model. Fig.~\ref{ClassicalGS} shows the classical ground state \cite{Misguich:2005} of the $J_{1}(=t^2/4U)-J_{2}(=(t^{\prime})^2/4U)$ Heisenberg model
 on the square lattice as a function of ${t^\prime/t}=\sqrt{J_{1}/J_{2}}$. Commensurate
 $(\pi,\pi)$ ordering is found for $t^{\prime}<\sqrt{\frac{1}{2}}t$ while for $t^{\prime}>\sqrt{\frac{1}{2}}t$ two independent AF sublattices appear as the ground state. Thermal or quantum fluctuations select a collinear phase\cite{Misguich:2005} with $(\pi,0)$ or $(0,\pi)$, and $(q_{x},\pi)$ and $(\pi,q_{y})$ for $t^{\prime}=\sqrt{\frac{1}{2}}$.
We can see in Figs.~\ref{f(t)} and ~\ref{Position-q} that at weak to intermediate coupling, Fermi surface effects as well as thermal and quantum fluctuations reduce the region  where $(\pi,\pi)$ commensurate ordering is found, relative to the classical strong coupling theory ($U>>t$).

  We now turn to the regions, illustrated in the lower part of Fig.1, where spin fluctuations can induce d-wave superconductivity. There are only five real irreducible representations of the tetragonal group that correspond to singlet pairing~\cite{Powell:2006}. Amongst these, only $B_{1g}$ ($d_{x^2-y^2}$) and $B_{2g}$ ($d_{xy}$) have an internal structure that can take advantage of spin fluctuations in the two-dimensional plane (see discussion section below). Incommensurate fluctuations could conceivably lead to triplet superconductivity in one of the five odd irreducible representations, a possibility we do not consider here.

  At $n=1$ and $t^\prime=0$, it was shown in Ref.~\onlinecite{Kyung:2003} that strong antiferromagnetic fluctuations create a pseudogap over the perfectly nested Fermi surface, suppressing the possibility of d-wave superconductivity.  By including frustration in the form of second-neighbor hopping, the pseudogap becomes limited to the hot spot regions~\cite{VilkShadow:1997}, allowing the $d_{x^2-y^2}$ superconducting state to appear. As the ordering wave vector becomes more incommensurate with increasing $t^\prime$, the conditions necessary for $d_{x^2-y^2}$ superconductivity are no-longer satisfied (see discussion section below). This explains the finite region over which $d_{x^2-y^2}$ appears in Fig.~\ref{Phase_Diagram}. Magnetic fluctuations near $(\pi,0)$ lead to $d_{xy}$ symmetry (see discussion section).   The area between the solid lines with circles (green) and the  dashed lines with
  triangles (blue) indicate respectively the $d_{{x^2}-y^2}$ and $d_{xy}$  superconducting regions.
  We can see that the window of d-wave superconductivity decreases upon decreasing $U$
  and vanishes at $(U=1.9,t^\prime=0.09)$ for $d_{x^2-y^2}$ and at $(U=3,t^\prime=1.85)$
  for the $d_{xy}$ state. When $U$ is not large enough in the presence of frustration, antiferromagnetic fluctuations are not large enough to lead to d-wave superconductivity. The large $U(\ge6)$ effect on the window where d-wave superconductivity exists cannot be captured  within TPSC since this approach is valid only from weak to
  moderate coupling.

   The region $t^\prime<0.5$ in the above phase diagram bears some similarity with that
   found by $T=0$ extrapolation of the fluctuation exchange approximation (FLEX)~\cite{Kondo:1999}.
   The value of ($U$, $t^\prime$) at which the $d_{{x^2}-y^2}$ superconducting
   state ceases to exist is around $(3.38,0.33)$ in that approach. The $d_{{x^2}-y^2}$ window increases rapidly with increasing $U$ and becomes approximately
   constant beyond $U=4.0$. (We are comparing the results of this paper for positive $t^\prime$ since they do not show the expected symmetry under change of sign of $t^\prime$ at $n=1$.) In the $T=0$ VCA study\cite{Andriy:2007} mentioned above, the $d_{x^2-y^2}$ state
    appears all the way down to $U=0$ and up to $t^\prime=1$. Our results are expected to be more accurate in the weak-coupling region. With TPSC we cannot address the question of $T=0$ homogeneous
   coexistence of superconductivity and antiferromagnetism (superconducting antiferromagnet).

\begin{figure}[tbp]
\begin{center}
\includegraphics[scale=0.4]{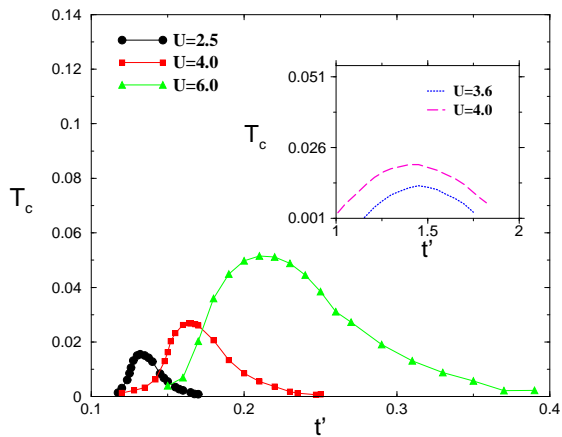}
\end{center}
\caption{(Color online) The $d_{x^2-y^2}$ superconducting critical temperature $T_c$ as a function of $t^\prime$ at $U=2.5$, $3$ and $4$ for $n=1$ . The inset shows the $d_{xy}$ superconducting critical temperature $T_c$ as a function of $t^\prime$ for $U=3.6$ and $4$.}
\label{Tc}
\end{figure}

The $d_{x^2-y^2}$ superconducting critical temperature $T_c$ as
a function of $t^\prime$ at $U=2.5$, $3$ and $4$ is depicted in Fig.~\ref{Tc}.
The inset shows $T_c(t^\prime)$ for the $d_{xy}$ case
at $U=3.6, 4$.
The maximum $T_c$ increases with increasing $U$, by contrast with what is found at strong coupling~\cite{Kotliar:1988,Senechal:2005,Kancharla:2007}.  $T_c$ for
the $d_{x^2-y^2}$ state as a function of $t^\prime$ at fixed U
shows a behaviour that is similar to that found earlier as a function of doping~\cite{Kyung:2003} for $t^\prime=0$. In fact, even the maximum value of $T_c$ that can be reached for given $U$ is comparable in the two cases. There are some similarities in the physics. The antiferromagnetic fluctuations need to be large enough and also at the correct wave vector to lead to superconductivity, but if they are too large with hot spots that cover the whole Fermi surface, superconductivity disappears.
In FLEX~\cite{Kondo:1999} there is no pseudogap but the $T=0$ extrapolation leads to antiferromagnetism destroying superconductivity at $t^\prime=0$.

We now study in turn how incommensuration, antiferromagnetic correlation length, double occupancy and Fermi surface effects influence d-wave superconductivity.

\begin{figure}[tbp]
\begin{center}
\includegraphics[width=7.5cm]{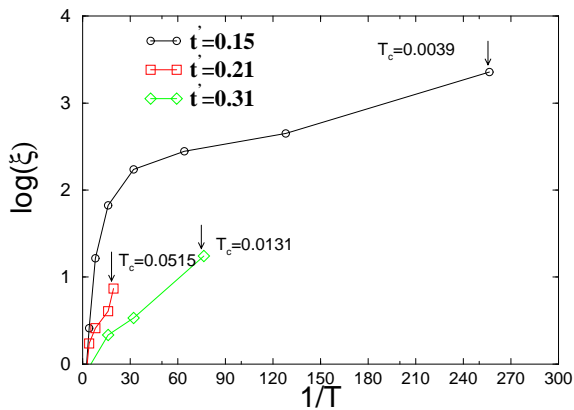}
\end{center}
\caption{(Color online) Logarithm base ten of the antiferromagnetic correlation length (in units of the lattice spacing) as a function of inverse temperature for three values of $t^\prime=0.15, 0.21, 0.31$ at $U=4$ for $n=1$. The value of $T_c$ for the corresponding $t^\prime$ is shown on the plot.}
\label{AFM_Length}
\end{figure}

Fig.~\ref{f(t)} illustrates at the crossover temperature $T_X$ the incommensurability of the wave vector where the maximum spin response occurs. The quantity $f(t^\prime)=\sqrt{(0.5-k_x^{max})^2+(0.5-k_y^{max})^2}$ is plotted as a function of $t^\prime$ at $U=2$, $4$ and $6$. Here, $k_x^{max}$ and $k_y^{max}$ are the $x$
and $y$ components of the wave vector where the spin response function is maximum. In $f(t^\prime)$, $k_x^{max}$ and $k_y^{max}$
are measured in units of $2\pi$ so that $0.5$ corresponds to $\pi$. The position
of the peak is at $(0.5,0.5)$ for small $t^\prime$ and approaches $(0.5,0)$
with increasing $t^\prime$. The peak position at $T_X$ does not
strongly depend on $U$ even though $T_X$ depends on $U$. To be more explicit, we show in Fig.~\ref{Position-q} how the wave vector moves in
the Brillouin zone. Since the spin correlation function has typically several maxima,
the highest peak determining the largest spin correlations sometimes suddenly jumps
from one place to another in momentum space, which clearly happens near $t^\prime=0.55$ and $0.7$. Since the Fermi surface topology changes near $t^\prime=0.71$, this is not surprising.

As can be seen by comparing Fig.~\ref{Position-q} and Fig.~\ref{Tc}, the maximum value of $T_c$ for $d_{{x^2}-y^2}$ superconductivity always occurs in the region where the antiferromagnetic fluctuations are commensurate at $(\pi,\pi)$, namely for $0<t^\prime<0.32$. Similarly, the maximum $T_c$ for $d_{xy}$ superconductivity occurs where magnetic fluctuations are nearly commensurate at $(0,\pi-\delta)$, although in the latter case, strict commensurability is never achieved. The maximum $T_c$ at $U=4$, for example, is smaller for $d_{xy}$ compared with $d_{{x^2}-y^2}$.



\begin{figure}[tbp]
\begin{center}
\includegraphics[scale=0.4]{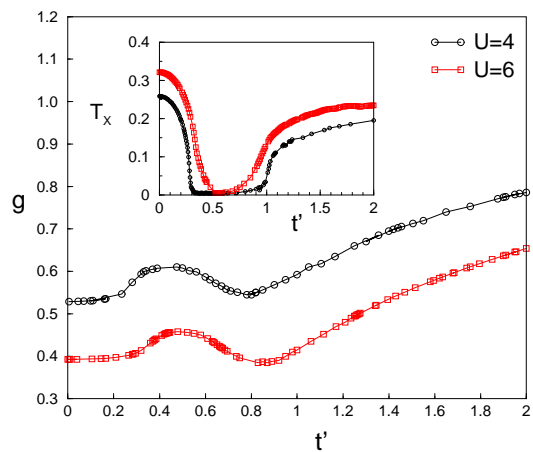}
\end{center}
\caption{(Color online) $g=g_{\uparrow\downarrow}(0)$ as a function of $t^\prime$ at $U=4$ and $6$ for $n=1$
at the crossover temperature $T_{X}$ . Inset shows $T_{X}$ as a function of
 $t^\prime$ at $U=4$ and $6$ for $n=1$. }
\label{DoubleOcc}
\end{figure}

The antiferromagnetic correlation length $\xi$ is a measure of how strong are the antiferromagnetic fluctuations.  Here, we find $\xi$ by calculating the curvature of the zero-frequency interacting spin susceptibility at the maximum. To assess the importance of this factor on superconductivity, the logarithm of $\xi$ is shown in Fig.~\ref{AFM_Length} for three different values of $t^\prime$ at $U=4$. An arrow indicates for what length we find a superconducting $T_c$. The antiferromagnetic correlation length at $T_c$ is smallest for the largest $T_c$. Near perfect nesting (smallest $t^\prime$), where pseudogap effects are important, superconductivity is hindered by the removal of states near zero energy~\cite{Kyung:2003}. To compensate, the antiferromagnetic correlation length has to become very large before superconductivity can set in, as we can see from $t^\prime=0.15$ on the figure. We can also compare at $T_c$ the thermal de Broglie wavelength $v_F/\pi T$ in units of the lattice spacing (and $k_B=1$, $\hbar=1$) with the value of $\xi$. In our units, the thermal length takes a value close to $1/T$ on the horizontal axis of Fig~\ref{AFM_Length}. It is only for the largest value $t^\prime=0.31$ that the superconducting $T_c$ occurs well above the renormalized classical regime, i.e. for the thermal length much larger than $\xi$. The important qualitative fact is that $T_c$ occurs below the renormalized classical regime $T_X$ when its occurrence is hindered by the pseudogap and that it can occur above $T_X$ otherwise. Near the maximum $T_c$, the two lengths are of the same order, in other words, for that $t^\prime=0.21$, $T_c$ and $T_X$ are of the same order. But in all cases $\xi$ is much larger than the lattice spacing.

For large $t^\prime$, where $d_{xy}$ superconductivity occurs, the antiferromagnetic correlation length is large, corresponding to the smaller $T_c$ obtained for that pairing symmetry (not shown).

\begin{figure}[tbp]
\begin{center}
\includegraphics[angle=-90,width=7cm]{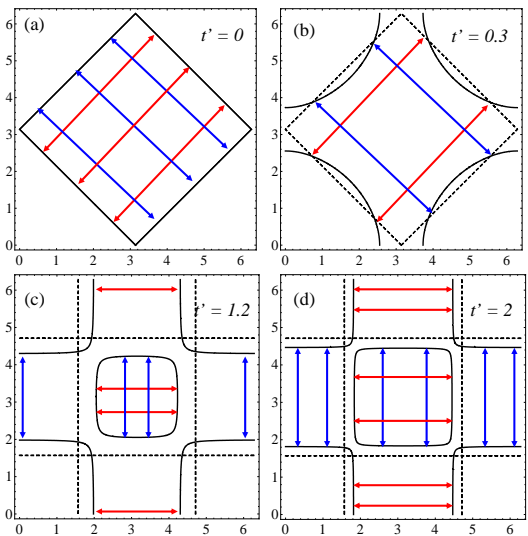}
\end{center}
\caption{(Color online) Non-interacting Fermi surfaces (black solid lines) at half-filling and wave vectors of the dominant spin fluctuations for four values of $t^\prime$ and $0<k_{x,y}<2\pi$. For negative values of $t^\prime$ the plots are identical if $-\pi<k_{x,y}<\pi$. The arrows represent the wave vectors that lead to the largest spin susceptibility. a) $t^\prime=0$ b) $t^\prime=0.3$. Dashed line is the antiferromagnetic Brillouin zone. c) $t^\prime=1.2$. d) $t^\prime=2.0$. In the last two plots, the dashed lines are separated by $(0,\pi)$ and $(\pi,0)$.}
\label{Fermi_surfaces}
\end{figure}

Returning to the factors influencing d-wave superconductivity, it is meaningful to ask how double occupancy in the underlying normal state influences the appearance of superconductivity. One might naively think that larger double occupancy would be favorable to superconductivity. A counter argument is that larger interaction strength, hence smaller double occupancy, is more favourable to d-wave superconductivity, hence should correspond to larger $T_c$. We show in Fig.~\ref{DoubleOcc} double occupancy normalized to its Hartree-Fock value, namely $g=g_{\uparrow\downarrow}(0)$, as a function of $t^\prime$ at $U=4$ and $6$ at the
crossover temperature $T_{X}$. The value of $T_{X}$  as a function of $t^\prime$ for the same values
of $U$ is shown in the inset. The maximum in double occupancy in the region $0.4 \leq t^\prime \leq 0.8$ corresponds to weaker spin fluctuations (as expected from the sum rule Eq.(5)) and vanishing $T_c$. This
is in agreement with what one can observe (yellow region) in Fig.~\ref{Phase_Diagram}. Comparing the overall values of $g=g_{\uparrow\downarrow}(0)$, we check, as expected, that double occupancy is smaller at larger $U$. The smaller double occupancy for larger $U$ corresponds to a larger value of the maximum $T_c(t^\prime)$ in Fig.~\ref{Tc}. Hence, from all of the above one might conclude that smaller double occupancy is better for d-wave superconductivity. However, for the $d_{{x^2}-y^2}$ case the maximum in $T_c$ as a function of $t^\prime$ occurs for both values of $U$ in a region where double occupancy is constant, independent of $t^\prime$. Hence, one cannot conclude that smaller double occupancy is more favorable to superconductivity. In fact the opposite trend is seen in the fact that the maximum $T_c$ at $U=4$ in the $d_{xy}$ case occurs in a region where double-occupancy is large. Even more striking is the fact that the value of double occupancy in the $d_{xy}$ case ($t^\prime>1$) is monotonically increasing with $t^\prime$ while $T_c$ is not. Clearly, double occupancy, a local quantity, is not directly correlated with the value of $T_c$ for order parameters with nodes at the origin.

We move to the influence of the shape of the Fermi surface on d-wave superconductivity. First of all, the shape of the Fermi surface influences the wave vector at which magnetic fluctuations are largest. Fig.~\ref{Fermi_surfaces} illustrates the fact that at $t^\prime=0$ perfect nesting occurs for $(\pi,\pi)$ and symmetry related vectors while for $t^\prime=2$ one has nearly perfect nesting at $(0,\pi-\delta)$. The change in Fermi surface topology occurs around $t^\prime=0.71$. There is no further change in topology between $t^\prime=0.71$ and $t^\prime=2$. The symmetry of the induced superconductivity is tied to the wave vectors of the magnetic fluctuations, namely $d_{{x^2}-y^2}$ for $(\pi,\pi)$ and $d_{xy}$ for $(0,\pi)$. No singlet superconductivity occurs in the vicinity of the change in Fermi surface topology because of the incommensurability and smallness of the fluctuations.

\begin{figure}[tbp]
\begin{center}
\includegraphics[width=8.5cm]{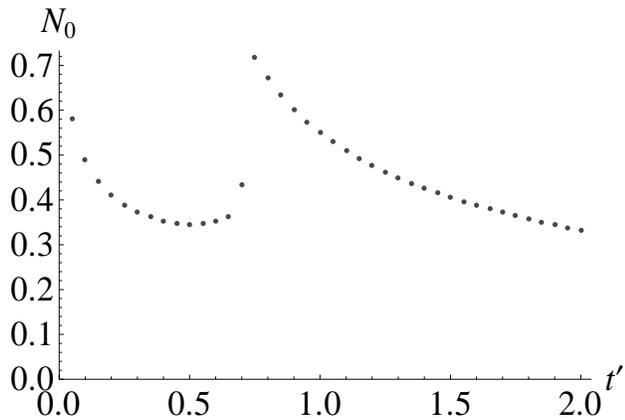}
\end{center}
\caption{(Color online) Non-interacting single particle density of states at the Fermi level $N_0$ as a function of $t^\prime$, calculated at half-filling with an energy resolution $0.02$. Spin degeneracy is included.}
\label{DOS}
\end{figure}

The single particle density of states of the non-interacting system is not a major factor in the occurrence of d-wave superconductivity. This is demonstrated in Fig.~\ref{DOS} that shows that there is no correlation with the maximum $T_c$ in Fig.~\ref{Tc} and peaks in the single particle density of states. There is a jump in density of states near $t^\prime=0.71$ where Fermi surface topology changes. The only case where the Fermi surface coincides with a van Hove singularity in the single-particle density of states is at $t^\prime=0$ and in that case there is no d-wave superconductivity, contrary to naive expectations. This is because the nesting allows strong antiferromagnetic fluctuations at $(\pi,\pi)$ to open a pseudogap over the whole Fermi surface. This is illustrated in Ref.~\onlinecite{Moukouri:2000,Kyung:2003}. The pseudogap removes density of states from the Fermi energy and is detrimental to d-wave superconductivity~\cite{Kyung:2003}. The same phenomenon is seen at $t^\prime=2$ where there is also nearly perfect nesting. In the presence of a finite $t^\prime$, the pseudogap is created only at hot spots and d-wave superconductivity can arise.


\section{discussion}

\begin{figure}[tbp]
\begin{center}
\includegraphics[angle=-90,width=8cm]{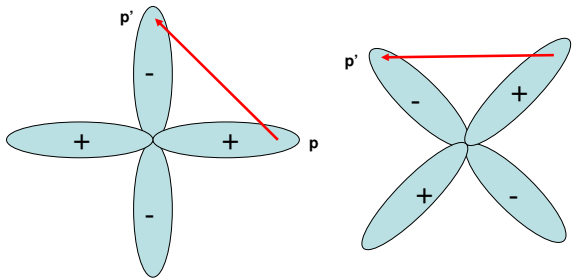}
\end{center}
\caption{(Color online) Illustration of the scattering by $(-\pi,\pi)$ fluctuations for $d_{x^2-y^2}$-wave superconductivity and by $(-\pi+\delta,0)$ for $d_{xy}$-wave superconductivity.}
\label{BCSorigin}
\end{figure}

Our estimate of $T_c$ from the normal state rests on the fact that at $T_c$ the vertex equation deduced from the Bethe-Salpeter equation becomes singular. In that case, the vertex equation has the same structure, symmetry-wise, as the BCS gap equation. One can thus understand very simply the relation between the symmetry of the d-wave order parameter and the underlying magnetic fluctuations. Normally, one would project the interaction potential in the different pairing channels~\cite{Beal-Monod:1986,Scalapino:1986,Miyake:1986}. A less rigorous but more intuitive approach follows Ref.~\onlinecite{Scalapino:1995} and starts from the BCS gap equation
\begin{equation}
\Delta_{\mathbf{p}}=-\frac{1}{2}\int\frac{d^{2}p^{\prime}}{\left(
2\pi\right)  ^{2}}V\left(  \mathbf{p-p}^{\prime}\right)  \frac{\Delta
_{\mathbf{p}^{\prime}}}{E_{\mathbf{p}^{\prime}}}\left(  1-2n\left(
E_{\mathbf{p}^{\prime}}\right)  \right)  .
\end{equation}
where $E_{\mathbf{p}^{\prime}}$ is the quasiparticle energy and $n(E_{\mathbf{p}^{\prime}})$ the Fermi function.  Even though the effective potential $V\left(\mathbf{p-p}^{\prime}\right)$ for singlet pairing due to spin fluctuations is always positive (repulsive), it can lead to an effective negative (attractive) potential for $d_{x^2-y^2}$-wave superconductivity if it is peaked at $(-\pi,\pi)$ and symmetry related wave vectors, as illustrated on the left of Fig.~\ref{BCSorigin}. Indeed, take a value $\mathbf{p}$ along the $x$ axis where the gap is positive. Then the largest contribution to the integral over $\mathbf{p'}$ will come from a region around $\mathbf{p'}$ located at $\pm 90^0$, where the gap is negative, such that the condition $\mathbf{p-p}^{\prime}=(-\pi,\pm \pi)$ is satisfied as closely as possible. The change in sign of the gap allows a solution of the gap equation even with $V\left(  \mathbf{p-p}^{\prime}\right)$ positive. (Clearly also, increasing the strength of the interaction will lead to a higher $T_c$.) The analogous argument explains why $(\pm \pi,0)$, $(0,\pm \pi)$ scattering favors $d_{xy}$-wave superconductivity, and why magnetic fluctuations that are peaked at incommensurate wave vectors are not favorable to any of the allowed singlet pairing symmetries on the square lattice. (In addition for a given $U$ the magnitude of fluctuations at incommensurate vectors is smaller than at commensurate vectors). While we have not studied the following question in detail, we note that incommensurate wave vectors may be favorable to pairing in triplet channels such as p-wave, ($E_u$)~\cite{Powell:2006}. There is however a change in sign of the effective interaction in the triplet channel~\cite{Beal-Monod:1986,Scalapino:1986,Miyake:1986}. That kind of mechanism may occur in the heavy fermion compound $CeCu_2Si_2$~\cite{Yuan:2003} for example.

According to Anderson~\cite{Anderson:2007}, that physics whereby
the pair state is orthogonal to the repulsive core of the Coulomb interaction was explained by Pitaevskii and Br\"uckner~\cite{Pitaevskii:1960,Brueckner:1960}. Note also that in our approach, we neglect the feedback of superconductivity on spin fluctuations~\cite{Leggett:1975}.

As discussed in the previous section, in the present case the non-interacting density of states has little influence on the value of $T_c$. Self-energy effects can strongly modify the effective density of states by creating a pseudogap that is detrimental to superconductivity. Self-energy effects are not so important in calculations that do not contain the pseudogap~\cite{Millis:1992}. We saw the suppression of superconductivity by the pseudogap at $t^\prime=0$ for example. That suppression is already apparent at high temperature in Quantum Monte Carlo calculations~\cite{Kyung:2003} and is reproduced by TPSC within a few percent in that regime. In apparent contradiction with the suppression of superconductivity by strong antiferromagnetic fluctuations in the presence of nesting, in quantum cluster approaches~\cite{Maier:2005,Maier_d:2005,Kancharla:2007} one does not see the suppression of d-wave superconductivity at half-filling for $t^\prime=0$ at weak to intermediate coupling (i.e. below the Mott transition), unless long-range antiferromagnetism is allowed. This is because for the small cluster sizes considered, the normal-state magnetic correlation length cannot become large enough to create a weak coupling pseudogap. Indeed, a weak coupling pseudogap can barely be seen in Ref.~\onlinecite{Moukouri:2001} for cluster sizes $4 \times 4$, even though they are larger than the $2 \times 2$ clusters in Ref.~\onlinecite{Kancharla:2007}. It is only at even larger cluster sizes that the weak coupling pseudogap appears in DCA calculations~\cite{Moukouri:2001}. The difference between weak and strong coupling pseudogaps has been discussed for example in Refs.~\onlinecite{Senechal:2004,Hankevych:2005}.

In electron-phonon mediated superconductivity, it has been found that the phonon frequencies near $7k_BT$ are the most efficient ones for pairing~\cite{carbotte:1986}. In the present case the wave vector of the magnetic fluctuations is also important. Although we have not studied the importance of the frequency dependence in detail, it is possible to make the following remarks. Take $d_{x^2-y^2}$ superconductivity. For small $t^\prime$, where superconductivity is reduced by self-energy effects coming from the pseudogap, the important frequencies are less than temperature in energy units because $T_c$ occurs in the renormalized classical regime. At larger $t^\prime$, where $T_c$ is reduced by incommensuration effects and deviations to nesting, the pseudogap occurs at lower temperature than $T_c$ hence finite frequencies larger than temperature are still important. Optimal $T_c$ occurs between these two cases.

\section{conclusion}

We have studied the conditions for the appearance of magnetic-fluctuation induced d-wave superconductivity in the half-filled Hubbard model in two dimensions. We have shown that at weak to intermediate coupling, the symmetry of the d-wave order parameter is determined by the wave vector of the magnetic fluctuations. Those that are near $(\pi,\pi)$ lead to $d_{x^2-y^2}$-wave ($B_{1g}$) superconductivity while those that are near $(0,\pi)$ induce $d_{xy}$-wave ($B_{2g}$) superconductivity. The dominant wave vector for magnetic fluctuations is determined by the shape of the Fermi surface so we find that $d_{x^2-y^2}$-wave superconductivity occurs for values of $t^\prime$ that are relatively small while $d_{xy}$-wave superconductivity occurs for $t^\prime>1$. For intermediate values of $t^\prime$ the magnetic fluctuations are smaller and incommensurate so no singlet superconductivity appears. The maximum value that $T_c$ can take as a function of $t^\prime$ increases with interaction strength. All of the above can easily be understood physically from simple BCS-like arguments. However, contrary to what is expected from BCS, the non-interacting single-particle density of states does not play a dominant role. With Fermi surfaces where hot spots can create a pseudogap on nearly all the Fermi surface (as in the $t^\prime=0$ case), self-energy effects hinder d-wave superconductivity, even though the strength of magnetic fluctuations can be very large. In that case, the large inelastic scattering rates are pair breaking and remove states near the Fermi level, decreasing the tendency to d-wave superconductivity. There is thus an optimal value of $t^\prime$ (frustration) for superconductivity. For $d_{x^2-y^2}$ superconductivity in underfrustrated systems (small $t^\prime$) $T_c$ occurs below the temperature $T_X$ where the crossover to the renormalized classical regime occurs. In other words, at $T_c$ the antiferromagnetic correlation length is much larger than the thermal de Broglie wave length. The opposite relationship between these lengths occurs for overfrustrated systems ($t^\prime$ larger than optimal) where $T_c$ is larger than $T_X$. The two temperatures are comparable for optimally frustrated systems. But in all cases, at $T_c$ the antiferromagnetic correlation length is larger than the lattice spacing.

Comparisons with experiments on heavy-fermion compounds, quasi one-dimensional organic metals and the question of the interplay of incommensurate magnetic fluctuations and triplet pairing is left for future work.

\section*{Acknowledgments}
A.-M.S.T. would like to thank C. Bourbonnais, J. Carbotte, F. Marsiglio, M. Norman and D.J. Scalapino for informative and stimulating discussions. Numerical calculations were performed on RQCHP computers and on a Sherbrooke's Elix cluster. The present
work was supported by NSERC (Canada), FQRNT (Qu\'{e}bec), CFI (Canada),
CIFAR, and the Tier I Canada Research chair Program (A.-M.S.T.).



\end{document}